\documentclass[aps,prb,twocolumn,showpacs]{revtex4}

\usepackage{graphicx}
\def  \bsig    {\mbox{\boldmath$\sigma$}}

\begin{document}

\title{Current-induced motion of a domain wall in magnetic nanowires}

\author{V. K. Dugaev$^{1,2,\, }$\cite{email}, V. R. Vieira$^1$, P. D. Sacramento$^1$,
J. Barna\'s$^{3}$, M. A. N. Ara\'ujo$^4$, and J. Berakdar$^5$}
\affiliation{$^1$Departamento de F\'isica and CFIF, Instituto Superior T\'ecnico,
Av. Rovisco Pais, 1049-001 Lisboa, Portugal\\
$^2$Frantsevich Institute for Problems of Materials Science,
National Academy of Sciences of Ukraine, Vilde 5, 58001 Chernovtsy, Ukraine\\
$^3$Department of Physics, Adam Mickiewicz University,
Umultowska~85, 61-614~Pozna\'n, Poland, and\\
Institute of Molecular Physics, Polish Academy of Sciences,
Smoluchowskiego~17, 60-179~Pozna\'n, Poland\\
$^4$Departamento de F\'isica, Universidade de \'Evora, P-7000 \'Evora, Portugal, and\\
Centro de F\'isica, Universidade do Minho, P-4710-057 Braga, Portugal\\
$^5$Max-Planck-Institut f\"ur Mikrostrukturphysik, Weinberg 2, 06120 Halle, Germany}

\date{\today }

\begin{abstract}
The dynamics of current-induced motion of a magnetic domain wall
in a quasi-one-dimensional ferromagnet with both easy-axis and
easy-plane anisotropy,  is studied. We pay a special attention to
the case of a sharp domain wall, and calculate the spin torque
created by the electric current. The torque has two components,
one of which is acting as a driving force for the motion of the
domain wall while the other distorts its shape, forcing thus the
magnetic moments to deviate from the easy plane.
\end{abstract}
\maketitle

\section{Introduction}

The increased interest in the dynamics of domain walls in magnetic
nanostructures is rooted mostly  in  perspective applications in
novel spintronic devices. Several recent experiments demonstrated
that the motion of domain walls can be effectively controlled by
means of an external magnetic field or an electric current.
\cite{grollier03,yamaguchi04,yamanouchi04,saltoh04,klaui05}

The theory of domain wall motion has been worked out   long time
ago.\cite{smith58,malozemoff79,berger92} Within  this theory, the
magnetic dynamics is usually described in the framework of a
classical ferromagnet assuming separation of the magnetic and
electronic degrees of freedom. In this way
 three- or two-dimensional ferromagnets are treated.

Present studies of the domain wall dynamics are devoted to the
issue of the current-induced magnetic motion in nanowires and
nanoconstrictions. In the presence of an electric current, the
domain wall can move due to a spin torque exerted on  the magnetic
system by the spin-polarized electron gas. In addition, (linear)
momentum  can be transferred directly to the domain wall upon
scattering of the charge carriers. The problem of calculating the
spin torque is of a prime importance for the theory of the
domain-wall motion, however  only  few recent works  address
 this
point
specifically.\cite{takagi96,li04,dugaev03,urazhdin04,barnes04}

The domain-wall motion can be described by Landau-Lifshitz
equations, which have a well-known static solution for the domain
wall.  Finding  the corresponding dynamical solution is however a
nontrivial task, particularly  in the presence of an external
force. The standard way is to describe the domain wall dynamics
using an approximate scheme, which is based on physically
reasonable arguments, but a strict mathematical justification is
missing. The simplest approach is to use an approximate solution
that assumes the moving wall to have exactly the same shape as
deduced from the static solution.  The rigorous conditions for the
 range of validity   of this approximation are however unclear.
To obtain reliable results  numerical simulations  are needed.
\cite{ohe05}

Most recently, new approaches have been put forward which address
\cite{takagi96,waintal03} the calculation of the torque as well as
the solution of the dynamical equations of motion of the domain
wall. Some aspects of these works are developed further and
partially revised in this paper.

Our treatment is concentrated on
the spin torque and  on the wall dynamics in a magnetic nanowire
with a  domain wall which is sharp on the length scale set by the
wave length of the relevant charge carriers. The approach is thus
more appropriate for the case of magnetic semiconductors with a
small Fermi momentum of carriers (electrons or holes).\cite{ruster03}

\section{Current-induced spin torque}

Here we consider the spin torque transferred by the electric
current from the spin polarized electron system to the domain
wall. The main goal of our calculation is the demonstration of the
presence of two components of the torque, which tend to rotate the
magnetic moment in different directions.

We adopt a one-dimensional model for the charge carriers with a
point-like interaction between the electron spin $\bsig $ and the
magnetic moment ${\bf M}(x)$ located at a  point $x$ along the
wire
\begin{equation}
\label{1}
H_{int}=g\, \bsig \cdot {\bf M}(x),
\end{equation}
where $g$ is the coupling constant. Here the one-dimensionality of
the electronic system means that we consider the electrons within
a wire with transversal dimensions smaller than the electron
wavelength $\lambda $, so that only the lowest electron subband is
relevant. Strictly speaking, the coupling of the moment to the
electron spin depends on the coordinates $y,z$ that characterize
the location of the moment within the wire, $g(y,z)\sim \left|
\psi _0(y,z)\right| ^2$, where $\psi _0(y,z)$ is the wave function
of the transverse motion of electrons in the lowest subband. For
simplicity, in the following we neglect this dependence,
substituting it by an averaged coupling, $g(y,z)\to g=A^{-1}\int
g(y,z)\, dy\, dz$, where $A$ is the cross section of the wire.

Considering the scattering of electrons from a magnetic moment
${\bf M}(x)$ we assume that the moment is frozen in the point $x$
on the scale of the characteristic times of the electron motion.
This assumption renders possible calculation of the torque as in
the case of a static domain wall. The calculated torque is then
utilized for the investigation of the domain wall dynamics. Our
assumption  relies on  an adiabatic approximation insofar as we
require the time scale for the motion of the magnetic subsystem to
be slow as compared to that of the electrons.

To calculate the torque in the case of a sharp domain wall, we
start from a model describing the one-dimensional scattering from
a localized moment in nonmagnetic and magnetic wires. Then we use
the results obtained for the simplified models, to calculate the
torque acting locally on the moments within the domain wall.

\subsection{Single magnetic moment in a nonmagnetic wire}

Let us consider first scattering of electrons in a nonmagnetic
one-dimensional system (nonmagnetic nanowire). The electrons are
scattered from a single frozen magnetic moment ${\bf M}_0$
situated at the point $x=0$, i.e., ${\bf M}(x)={\bf M}_0\, \delta
(x)$. Here we denote the coordinate along the wire as $x$. In the
absence of the spin-orbit interaction, it is more convenient to
use a different coordinate system $(x^\prime ,y^\prime ,z^\prime
)$ for the spin space. We calculate the total scattering amplitude
(beyond the Born approximation) of an electron with an arbitrary
spin polarization, coming from $x=-\infty$, and elastically
scattered into the states with different spin polarizations.

Assuming the quantization axis $z^\prime $ along the moment ${\bf
M}_0$, we can write the spinor wave function of electrons as
\begin{eqnarray}
\label{2}
 \psi (x)=\left\{
 \begin{array}{c}
 e^{ikx} \left( \begin{array}{c} a\\ b\end{array}\right)
+e^{-ikx} \left( \begin{array}{c} ra\\
r^*b\end{array}\right) , \hskip0.5cm x<0, \\
\hspace*{-2.3cm} e^{ikx} \left( \begin{array}{c} ta\\
t^*b\end{array}\right) ,
\hskip0.5cm x>0, \\
\end{array}\right.
\end{eqnarray}
where the coefficients $a$ and $b$ correspond
to an arbitrary spin polarization in the incident electron wave,
$r$ and $t$ are, respectively, the reflection and the
transmission coefficients in the spin up channel
\begin{equation}
\label{3}
r=-\frac{i\alpha }{1+i\alpha }\; ,\hskip0.5cm
t=\frac1{1+i\alpha }\; ,
\end{equation}
$\alpha =gM_0m/k\hbar ^2$, and $M_0$ is the magnitude of the localized
magnetic moment.

Using Eqs. (2) and (3) we can calculate the spin density in the
wire, associated with the wave function (2)
\begin{equation}
\label{4}
S_\mu (x)=\psi ^\dag (x)\, \sigma _\mu \, \psi (x).
\end{equation}
$S_\mu (x)$
stands for  the oscillating spin density created by  the electrons reflected
backwards from the localized moment, $x<0$. The spin density components are
\begin{eqnarray}
\label{5}
S_{x^\prime }(x<0)=s_{x^\prime }\; \frac{1+\alpha ^2+2\alpha ^4}{(1+\alpha ^2)^2}
+s_{y^\prime }\; \frac{2\alpha ^3}{(1+\alpha ^2)^2}
\nonumber \\
-2\cos \, (2kx)\left( s_{x^\prime }\; \frac{\alpha ^2}{1+\alpha ^2}
+s_{y^\prime }\; \frac{\alpha }{1+\alpha ^2}\right) ,
\end{eqnarray}
\begin{eqnarray}
\label{6}
S_{y^\prime }(x<0)=s_{y^\prime }\; \frac{1+\alpha ^2+2\alpha ^4}{(1+\alpha ^2)^2}
-s_{x^\prime }\; \frac{2\alpha ^3}{(1+\alpha ^2)^2}
\nonumber \\
+2\cos \, (2kx)\left( -s_{y^\prime }\; \frac{\alpha ^2}{1+\alpha ^2}
+s_{x^\prime }\; \frac{\alpha }{1+\alpha ^2}\right) ,
\end{eqnarray}
\begin{eqnarray}
\label{7}
S_{z^\prime }(x<0)=s_{z^\prime }\left( \frac{1+2\alpha ^2}{1+\alpha ^2}
-\cos \, (2kx)\; \frac{2\alpha ^2}{1+\alpha ^2}\right)
\nonumber \\
-\sin \, (2kx)\; \frac{2\alpha }{1+\alpha ^2}\, ,
\end{eqnarray}
where $s_\mu $ is the unit vector determining the spin polarization in the
incident wave.
Similarly, we obtain for $x>0$
\begin{equation}
\label{8}
S_{x^\prime }(x>0)=s_{x^\prime }\; \frac{1-\alpha ^2}{(1+\alpha ^2)^2}
-s_{y^\prime }\; \frac{2\alpha }{(1+\alpha ^2)^2}\, ,
\end{equation}
\begin{equation}
\label{9}
S_{y^\prime }(x>0)=s_{y^\prime }\; \frac{1-\alpha ^2}{(1+\alpha ^2)^2}
+s_{x^\prime }\; \frac{2\alpha }{(1+\alpha ^2)^2}\, ,
\end{equation}
\begin{equation}
\label{10}
S_{z^\prime }(x>0)=s_{z^\prime }\; \frac1{1+\alpha ^2}\, .
\end{equation}
As follows from Eqs.~(5) to (10), the spin density induced by the
spin-polarized wave incoming from $x=-\infty $, oscillates with
the period $\pi /k$ at $x<0$, and is constant for $x>0$.

The spin current is defined as
\begin{equation}
\label{11}
j^s_\mu (x)
=\frac{i\hbar }{2m}\left[
(\nabla _{x}\psi ^\dag (x))\, \sigma _\mu \, \psi (x)
-\psi ^\dag (x)\, \sigma _\mu \, \nabla _{x}\psi (x)
\right] ,
\end{equation}
($\mu =x^\prime ,y^\prime ,z^\prime $) and it can be also
calculated using Eqs.~(2) and (3) for $x<0$ and $x>0$,
respectively. Then we find that the spin current is constant
for $x<0$ and $x>0$, with a jump of $x^\prime $ and $y^\prime $
components at $x=0$.

We can calculate the spin torque acting on the moment ${\bf M}_0$
as the transferred spin current at the point $x=0$
\begin{equation}
\label{12}
T_\mu=j^s_\mu (-\delta )-j^s_\mu (+\delta ).
\end{equation}
Using Eqs.~(2),(3),(11) and (12) we obtain
\begin{equation}
\label{13}
T_{x^\prime }=\frac{j_0}{e}\left[
s_{x^\prime }\, \frac{4\alpha ^2}{1+\alpha ^2}
+s_{y^\prime }\, \frac{2\alpha (1-\alpha ^2)}{1+\alpha ^2}\right] ,
\end{equation}
\begin{equation}
\label{14}
T_{y^\prime }=\frac{j_0}{e}\left[
s_{y^\prime }\, \frac{4\alpha ^2}{1+\alpha ^2}
-s_{x^\prime }\, \frac{2\alpha \, (1-\alpha ^2)}{1+\alpha ^2}\right] ,
\end{equation}
and $T_{z^\prime }=0$, where $e$ is the electron charge ($e<0$), $j_0$ is
the electric current
\begin{equation}
\label{15}
j_0=\frac{ie\hbar }{2m}\left[
(\nabla _{x}\psi ^\dag (x))\, \psi (x)
-\psi ^\dag (x)\, \nabla _{x}\psi (x)\right]
=\frac{ev}{1+\alpha ^2}\;
\end{equation}
and $v=\hbar k/m$ is the velocity. Note that we found the
components of torque (13), (14) in the coordinate system related
to the moment ${\bf M}_0$, so that $x^\prime $ and $y^\prime $
axes are perpendicular to the vector ${\bf M}_0$, and they are not
related in any way to the direction of the current $j_0$.

Using Eqs.~(5)-(10) one can show that exactly the same values for
the torque components (13),(14) are obtained by utilizing the
relation
\begin{equation}
\label{16}
T_\mu =-\frac{gM_0}{\hbar }\;
\epsilon _{\mu\nu\lambda}\, n_\nu \, S_\lambda (0),
\end{equation}
which follows from the equation of motion of the magnetic moment
${\bf M}_0$, where ${\bf n}$ is the unit vector along ${\bf M}_0$,
and $\epsilon _{\mu\nu\lambda}$ is the unit antisymmetric tensor.

One can also present the result for the torque (13),(14) in a
form, which is more appropriate for an arbitrary coordinate system
for the electron spin (not necessarily with the axis $z$ along
${\bf M}_0$)
\begin{equation}
\label{17}
T_\mu  =\frac{j_0}{e}\left[
\eta \left( \delta _{\mu \nu }-n_\mu n_\nu \right) s_\nu
+\zeta \, \epsilon _{\mu\nu\lambda}\, n_\nu \, s_\lambda
\right] ,
\end{equation}
where
\begin{equation}
\label{18}
\eta =\frac{4\alpha ^2}{1+\alpha ^2}\; ,\hskip0.5cm
\zeta =-\frac{2\alpha \, (1-\alpha ^2)}{1+\alpha ^2}\; .
\end{equation}

As we see from Eqs.~(17) and (18), there are two components of the
torque -- both transverse to the localized moment. Apart from
this, one of them tends to align the moment along the direction of
the spin polarization of the incoming electrons, whereas the other
one is perpendicular to the spin polarization of the incident
wave. Note that in the Born approximation for the scattering
amplitude, which is valid for $\alpha \ll 1$, only the second term
in (17) survives. It rotates the moment ${\bf M}_0$ to the
direction perpendicular to vector ${\bf s}$ (and also to  ${\bf
n}$).

Thus, the spin torque acting on a single magnetic moment can be
found as a change of the spin current due to scattering from the
localized moment. The same result is obtained from the calculation
of the interaction of accumulated spin with the localized moment.
In the following, to calculate the torque in the domain wall, we
will use both methods but the second one (coupling to the
accumulated spin) is more convenient in case of a sharp domain wall.

We assume that in the case of a smooth domain wall, it can be more
convenient to solve the problem considering it as a propagation of
a spin-polarized wave with subsequent scattering from different
magnetic moments. Correspondingly, one can calculate the torque as
a divergence of the spin current.

\subsection{Scattering from a single magnetic moment in a magnetic wire}

Now we calculate the torque in the case of a magnetic wire with
the magnetization ${\bf M}$ being oriented along the axis $x$ for
$x<0$ (left to the wall) and in the opposite direction for $x>0$
(right to the wall). Here we assume the spin coordinate system
$(x^\prime ,y^\prime ,z^\prime )$ to coincide with the the
$(x,y,z)$ one. Like in the previous problem, we introduce an
additional frozen magnetic moment ${\bf M}_0=M_0\left( n_x,\,
n_y,\, 0\right) $ located at the point $x=0$. For definiteness,
let the vector ${\bf M}_0$ lie in the $x-y$ plane.

The relevant Hamiltonian has the form
\begin{eqnarray}
\label{19}
H=-\frac{\hbar ^2}{2m}\frac{d^2}{dx^2}+gM\, \sigma _x\, {\rm sgn}\, (x)
+gM_0\, {\bf n}\cdot \bsig\, \delta (x),
\end{eqnarray}
First, we consider the torque created by a single spin-polarized
wave (with the spin polarization along the axis $x$ labelled as
"$\uparrow $") coming from the left. We choose the quantization
axis along the axis $z$. Then the wave function containing the
reflected and transmitted waves of opposite polarization is
\begin{eqnarray}
\label{20}
\psi _\uparrow (x)=
\left\{
\begin{array}{c}
 \frac{e^{ik_\uparrow x}+r_\uparrow \,
e^{-ik_\uparrow x}}{\sqrt{2}} \left( \begin{array}{c} 1\\
1\end{array}\right) +\frac{r_{\uparrow f}\, e^{-ik_\downarrow
x}}{\sqrt{2}} \left( \begin{array}{c} 1\\ -1\end{array}\right),\\
\hspace*{3cm}\quad \mbox{for} \quad x<0, \\ \\
   \frac{t_\uparrow \, e^{ik_\downarrow x}}{\sqrt{2}}
\left( \begin{array}{c} 1\\ 1\end{array}\right) +\frac{t_{\uparrow
f}\, e^{ik_\uparrow x}}{\sqrt{2}}
\left( \begin{array}{c} 1\\ -1\end{array}\right),\\ \hspace*{3cm}\quad \mbox{for} \quad x>0,\\
\end{array}\right.
\end{eqnarray}
where $k_{\uparrow ,\downarrow }=[2m(\varepsilon \mp
gM)]^{1/2}/\hbar $, and $\varepsilon $ is the energy. Note that
the spin-up electrons are the spin-minority ones, while the
spin-down electrons are the spin-majority ones.

Using the continuity of wave function at $x=0$ and the condition
resulting from the integration of Schr\"odinger equation in the
vicinity of domain wall,
\begin{equation}
\label{21}
-\frac{\hbar ^2}{2m}\left(
\left. \frac{d\psi }{dx}\right| _{+\delta }
-\left. \frac{d\psi }{dx}\right| _{-\delta }\right)
+gM_0\left( n_x\sigma _x+n_y\sigma _y\right) \psi (0)=0,
\end{equation}
we find the transmission coefficients for the spin-up polarized wave
\begin{equation}
\label{22}
t_\uparrow =\frac{2k_\uparrow \left( k_\uparrow +k_\downarrow -ig_on_x\right) }
{\left( k_\uparrow +k_\downarrow \right) ^2+g_0^2}\; ,
\end{equation}
\begin{equation}
\label{23}
t_{\uparrow f}=-\frac{2g_on_yk_\uparrow }
{\left( k_\uparrow +k_\downarrow \right) ^2+g_0^2}\; ,
\end{equation}
and the reflection factors $r_\uparrow =t_\uparrow -1$, $r_{\uparrow f}=t_{\uparrow f}$,
where we denote $g_0=2gmM_0/\hbar ^2$.

Using Eqs.~(11) and (20), we can calculate the components of the spin current
induced by the incoming spin-up wave
\begin{eqnarray}
\label{24}
j^s_{\uparrow x}(x)=\left\{
\begin{array}{c}
 v_\uparrow \, (1-|r_\uparrow |^2)+v_\downarrow \, |r_{\uparrow f}|^2,
\hskip0.5cm x<0, \\
\\
 v_\downarrow \, |t_\uparrow |^2-v_\uparrow \, |t_{\uparrow f}|^2,
\hskip0.5cm x>0,  \\
\end{array}
\right.
\end{eqnarray}
\begin{eqnarray}
\label{25}
j^s_{\uparrow y}(x)=
\left\{
\begin{array}{c}
 t_{_\uparrow f}\, {\rm Im}\left[ v_\uparrow \,
(e^{ik_+x}-r_\uparrow \, e^{-ik_-x})\hskip1.5cm \right.
\\ \left. +v_\downarrow \, (e^{-ik_+x}+r_\uparrow ^*\,
e^{ik_-x})\right] , \hskip0.5cm x<0,  \\
\\
 t_{_\uparrow f}\, {\rm Im}\left[
-v_\uparrow \, t_\uparrow ^*\, e^{ik_-x} +v_\downarrow \,
t_\uparrow \, e^{-ik_-x}\right] ,
\; x>0 ,\\
\end{array}
\right.
\end{eqnarray}
\begin{eqnarray}
\label{26}
j^s_{\uparrow z}(x)=
\left\{
\begin{array}{c}
  t_{\uparrow f}\, {\rm Re}\left[ v_\uparrow \,
(e^{ik_+x}-r_\uparrow \, e^{-ik_-x})\hskip1.5cm \right.
\\ \left. -v_\downarrow \, (e^{-ik_+x}+r_\uparrow ^*\,
e^{ik_-x})\right] , \hskip0.3cm x<0, \\
\\
t_{_\uparrow f}\, {\rm Re}\left[ v_\uparrow \, t_\uparrow ^*\,
e^{ik_-x} +v_\downarrow \, t_\uparrow \, e^{-ik_-x}\right] ,
\hskip0.2cm x>0,  \\
\end{array}
\right.
\end{eqnarray}
where $k_{\pm}=k_\uparrow \pm k_\downarrow $ and
$v_{\uparrow ,\downarrow }=\hbar k_{\uparrow ,\downarrow }/m$.

Hence, the transverse components of the spin currents,
$j^s_{\uparrow y}(x)$ and $j^s_{\uparrow z}(x)$, are nonzero for
$x<0$ and for $x>0$. As we see from Eqs.~(24)-(26), the transverse
components of the spin current are oscillating functions of $x$.
The nonconservation of spin current in the magnetic wire is related to
indirect magnetic interactions accompanying the inhomogeneous
distribution of the spin density.
In the nonmagnetic case, corresponding to the limit of $k_-\to 0$,
it reduces to the conservation of spin current at $x<0$ and $x>0$, as we
found in the previous section.

The calculation of the corresponding spin transfer (12) in this model
gives us the following expressions for the components of the torque
\begin{equation}
\label{27}
T_{\uparrow x}=2v_\uparrow \, {\rm Re}\, t_\uparrow
+\left( v_\uparrow +v_\downarrow \right) \left( |t_{\uparrow
f}|^2-|t_\uparrow |^2\right) ,
\end{equation}
\begin{equation}
\label{28}
T_{\uparrow y}=-2t_{\uparrow f}\left( v_\uparrow
+v_\downarrow \right) \, {\rm Im}\, t_\uparrow ,
\end{equation}
\begin{equation}
\label{29}
T_{\uparrow z}=2t_{\uparrow f}\left[ v_\uparrow -\left(
v_\uparrow +v_\downarrow \right) \, {\rm Re}\, t_\uparrow \right] .
\end{equation}

As in the case of a nonmagnetic wire with a localized moment, the
torque ${\bf T}_\uparrow $ can be  calculated using Eq.~(16) with
the spin density ${\bf S}_\uparrow (0)$, where the index
"$\uparrow $" indicates the accumulated spin associated with the
incident wave of spin-up polarization. Using (4) and (20) we can
calculate the spin density ${\bf S}_\uparrow (x)$ as in the
previous section. The spin density oscillates  for both $x<0$ and
$x>0$,
 with the periods  $2\pi /k_+$ and $2\pi /\left| k_-\right| $
as for the spin current in Eqs.~(24)-(26).

With our choice of the coordinate system, when the vector ${\bf M}_0$ lies
in the $x-y$ plane, the transverse components
of the torque acting on the moment ${\bf M}_0$ are
\begin{equation}
\label{30}
T_{\uparrow \perp }=-n_y\, T_{\uparrow x}+n_x\, T_{\uparrow y}\, ,
\end{equation}
rotating the moment in $x-y$ plane, and $T_{\uparrow z}$, rotating
it in out-of-plane direction. The label $\perp $ in (30) means the
projection of torque on the direction perpendicular to the moment
${\bf M}_0$ in the $x-y$ plane.

Similarly, we can consider the scattering of electron with the
incident spin polarization opposite to the $x$-axis (labelled as
$\downarrow $). The corresponding scattering state has the form of
Eq.~(20) with $k_\uparrow \leftrightarrow k_\downarrow $ and
interchanged spin states, and the expressions for transmission
coefficients are
\begin{equation}
\label{31}
t_\downarrow = \frac{2k_\downarrow \left( k_\uparrow +k_\downarrow
+ig_on_x\right) } {\left( k_\uparrow +k_\downarrow \right)
^2+g_0^2}\; ,
\end{equation}
\begin{equation}
\label{32}
t_{\downarrow f}=\frac{2g_on_yk_\downarrow }
{\left( k_\uparrow +k_\downarrow \right) ^2+g_0^2}\; ,
\end{equation}
where the change $g_0 \to -g_0$ is equivalent to the flip of momentum ${\bf M}_0$.
The components of spin current $j^s_{\downarrow \mu }$ have
the following form
\begin{eqnarray}
\label{33}
j^s_{\downarrow x}(x)=
\left\{
\begin{array}{c}
 -v_\downarrow \, (1-|r_\downarrow |^2)-v_\downarrow \,
|r_{\downarrow f}|^2, \hskip0.5cm x<0,  \\ \\
 -v_\uparrow \, |t_\downarrow |^2+v_\downarrow \, |t_{\downarrow f}|^2,
\hskip0.5cm x>0 , \\
\end{array}
\right.
\end{eqnarray}
\begin{eqnarray}
\label{34}
j^s_{\downarrow y}(x)=
\left\{
\begin{array}{c}
 -t_{\downarrow f}\, {\rm Im}\left[ v_\downarrow \,
(e^{ik_+x}-r_\downarrow \, e^{-ik_-x})\hskip1.5cm \right.
 \\ \left. +v_\uparrow \, (e^{-ik_+x}+r_\downarrow ^*\,
e^{ik_-x})\right] , \hskip0.2cm x<0,
 \\ \\
  t_{_\downarrow f}\, {\rm Im}\left[ v_\downarrow \,
t_\downarrow ^*\, e^{ik_-x} -v_\uparrow \, t_\downarrow \,
e^{-ik_-x}\right]  \hskip0.2cm x>0 ,\\
\end{array}
\right.
\end{eqnarray}
\begin{eqnarray}
\label{35}
j^s_{\downarrow z}(x)=
\left\{
\begin{array}{c}
t_{\downarrow f}\, {\rm Re}\left[
v_\downarrow \, (e^{ik_+x}-r_\downarrow \, e^{-ik_-x})\hskip1.5cm
\right. \\ \left. -v_\uparrow \,
(e^{-ik_+x}+r_\downarrow ^*\, e^{ik_-x})\right] , \hskip0.2cm x<0,
 \\ \\
t_{\downarrow f}\, {\rm Im}\left[ v_\downarrow \, t_\downarrow
^*\, e^{ik_-x} +v_\uparrow \, t_\downarrow \,
e^{-ik_-x}\right] , \hskip0.2cm x>0 .  \\
\end{array}
\right.
\end{eqnarray}

The torque components (27)-(29) can also be calculated from
Eq.(16), taking into account the spin accumulation at $x=0$,
\begin{eqnarray}
\label{36}
S_{\uparrow ,\downarrow \, x}(0) &=&
\frac{4k_{\uparrow ,\downarrow
}^2\, [(k_\uparrow +k_\downarrow )^2 +g_0^2\,
(n_x^2-n_y^2)]} {[(k_\uparrow +k_\downarrow )^2+g_0^2]^2}\; ,\\
%\end{equation}
%\begin{equation}
%
S_{\uparrow ,\downarrow \, y}(0) &=&
\frac{8g_0^2\, k_{\uparrow
,\downarrow }^2\, n_xn_y} {[(k_\uparrow +k_\downarrow
)^2+g_0^2]^2}\; ,\label{37}\\
%\end{equation}
%\begin{equation}
%
 S_{\uparrow
,\downarrow \, z}(0) &=&\mp \frac{8g_0\, k_{\uparrow ,\downarrow
}^2(k_\uparrow +k_\downarrow )\, n_y} {[(k_\uparrow +k_\downarrow
)^2+g_0^2]^2}\; . \label{38}
\end{eqnarray}
The above formulas will be used later to calculate the torque
exerted on a domain wall.

 In the case of a 100\% polarized electron gas, only the
components of the spin current calculated according to
Eqs.~(33)-(35) corresponding to the majority electrons, are
relevant. The transition to this case implies the substitutions
$k_\uparrow \to i\kappa _\uparrow $.

In this section we calculated the wave functions of electrons in
the magnetic wire with a sharp domain wall and a localized
magnetic moment. The eigenfunctions of the Hamiltonian (19)
correspond to the spin-polarized incoming waves (spin up and
down). Obviously, an arbitrary-polarized incoming wave is not the
eigenfunction of the Hamiltonian. Nevertheless, we can still
consider the scattering of electron waves with different spin
polarizations. For example, such a state can be created by means
of an injection from the tip, and its lifetime $\tau $ can be long
enough on the scale of   the characteristic time of the domain
wall motion. In this case, a superposition of states with
different incoming spin-polarized waves can be used to calculate
the torque.

\subsection{Magnetic wire with a thin domain wall}

In the case of a thin metallic wire, when $k_{F\uparrow ,\downarrow}d \ll 1$
($d$ is the wire diameter and $k_{F\uparrow ,\downarrow }$ are the Fermi
momenta of minority and majority electrons, respectively), we can assume
that only one quantization level is filled with electrons.

Let us consider a magnetic wire with a single domain wall, with
the magnetization ${\bf M}$ being directed along the $x$ axis for
$x<-w$ and along the opposite direction for $x>w$. Here  $2w$ is
the domain wall width, and we chose the spin coordinate system
like in the previous section, with the $x$ axis along the wire.

Upon applying  a small voltage an electric current flows in the
wire. For definiteness, we assume the current to flow in the
direction opposite to the  $x$ axis direction (i.e., the electron
flux is oriented along $x$). If the only imperfection in the wire
is the  presence of the domain wall, one can assume  a jump
$\Delta \phi $  in the potential at the wall, and both the charge
and the spin currents can be calculated as integrals over energies
in the interval between $\varepsilon _{FR}$ and $\varepsilon
_{FL}=\varepsilon _{FR}+e\, \Delta \phi $, where $\varepsilon
_{FL}$ and $\varepsilon _{FR}$ are  the Fermi levels on the left
and right sides. In the limit of small voltage, $\vert e\, \Delta
\phi\vert \ll \varepsilon _F$, the transport is linear and is
associated with electrons at the Fermi level.

\begin{figure}
\hspace*{-1cm}
\includegraphics[scale=0.6]{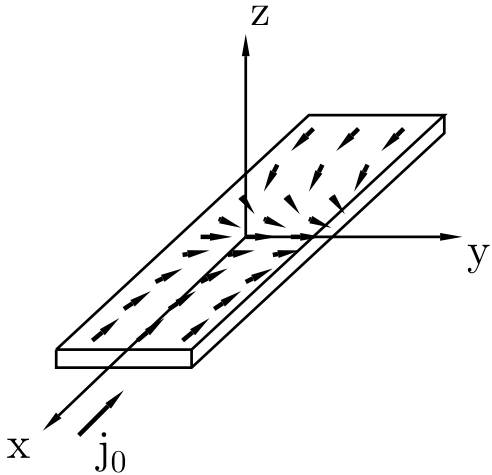}
\caption{Schematic picture of the domain wall.}
\end{figure}

We assume the electrons approaching  the domain wall from the left
are spin-polarized according to the magnetization direction in the
left part of the wire. The incoming electrons are scattered from a
large number of magnetic moments in the domain wall. We consider
this scattering using the point interaction of electron with each
of the localized moments. This corresponds to the picture with an
array of well-separated magnetic moments like in the case of
magnetic semiconductor doped with magnetic impurities. We assume
that there is a large number of magnetic atoms with different
orientation of moments within the domain wall. Accordingly, the
electron transmitted through the wall is multiply scattered from
many magnetic moments.

Thus, to calculate the transmission of electrons through the
domain wall, we should take the perturbation created by the total
magnetic moment $\widetilde{\bf M}(x)=\sum _i{\bf M}_i\, \delta
(x-x_i)$, where ${\bf M}_i$ is the localized moment at the point
$x=x_i$, and all of the moments ${\bf M}_i$ are located within a
region of the domain wall width, $|x_i|<w$, which in turn is
assumed to be small as compared to the wavelength of electrons,
$k_{F\uparrow, \downarrow }w\ll 1$.

The scattering from the total moment $\widetilde{\bf M}(x)$ located
within a region much smaller than the electron wavelength can be
described using the model of a spin-dependent delta-function
potential.\cite{dugaev03} Then, in the limit of small voltage, we
obtain the current
\begin{equation}
\label{39}
j_0\simeq \frac{e^2\, \Delta \phi }{2\pi \hbar }
\left( |\tilde{t}_{\uparrow f}| ^2
+\frac{v_\downarrow }{v_\uparrow }\; |\tilde{t}_\uparrow |^2
+|\tilde{t}_{\downarrow f}| ^2
+\frac{v_\uparrow }{v_\downarrow }\; |\tilde{t}_\downarrow |^2\right) ,
\end{equation}
where the tilde means the transmission coefficients for the
scattering of electrons from an effective moment\cite{dugaev03}
${\bf M}_{eff}\simeq \int _{-w}^{+w}\widetilde{\bf M}(x)\, dx$. It
is the B\"uttiker-Landauer formula for conductivity, which can be
obtained in the linear response approximation, using the basis of
scattering states. There are two contributions in Eq.~(39) related
to the incoming waves with spin up and spin down polarizations,
and with the corresponding momenta at the Fermi surface,
$k_{\uparrow ,\downarrow }\equiv k_{F\uparrow ,\downarrow }$.

In the case the domain wall possesses a  magnetization profile as
depicted in Fig.~1, the effective moment ${\bf M}_{eff}$ is
oriented along the $y$ axis. The transmission coefficients
$\tilde{t}_{\uparrow }$, $\tilde{t}_{\uparrow f}$ and
$\tilde{t}_{\downarrow }$, $\tilde{t}_{\downarrow f}$ have the
form of Eqs.~(22), (23) and (31), (32), respectively, with
$n_x=0$, $n_y=1$, and the substitution $g_0\to \tilde{g}_0\equiv
2mgM_{eff}/\hbar ^2$. We can relate the magnitude of $M_{eff}$ to
the continuous magnetic profile within the wall, $M_{eff}\simeq
\int _{-w}^wM_y(x)\, dx$.

The spin current can be also calculated in linear response
approximation using the scattering states.\cite{dugaev03} It
includes the sum of partial spin currents
\begin{equation}
\label{40}
{\bf j}^s(x)=\frac{e\, \Delta \phi }{2\pi \hbar }
\left( \frac{\tilde{\bf j}^s_\uparrow (x)}{v_\uparrow }
+\frac{\tilde{\bf j}^s_\downarrow (x)}{v_\downarrow }\right) ,
\end{equation}
where the components of $\tilde{\bf j}^s_{\uparrow ,\downarrow }$
can be found using Eqs.~(24)-(26) and (33)-(35) with the
substitution $t_{\uparrow ,\downarrow },\; t_{\uparrow ,\downarrow
f}\to \tilde{t}_{\uparrow ,\downarrow },\; \tilde{t}_{\uparrow
,\downarrow f}$. The appearance of $v_\uparrow $ and $v_\downarrow
$ in the denominators of Eq.~(40) is related to the 1D density of
states for spin up and down electrons. The components of the spin
current perpendicular to the $x$ axis, are oscillating functions.
As we see from Eqs.~(24)-(26) and (33)-(35), the wavelength of
oscillations is determined by the inverse momentum at the Fermi
level. Hence, the wavelength of oscillation of the transverse
component of the spin current is much larger than the domain wall
width.

It is worth
noting that in three-dimensional systems, the transverse component of the spin
current decays due to the integration over momentum in the DW plane.
In metallic ferromagnets, the decay is very fast due to the large Fermi
momentum of electrons.
However, there is a nonvanishing spin transfer for the transverse component
in the 3D case, too.

We can also calculate the accumulated spin density induced by the
external current $j_0$. It can be found as the expectation value
of the spin $\sigma _\mu $ in the scattering state of the incoming
electrons, integrated over all energies between $\varepsilon _F$
and $\varepsilon _F+e\, \Delta \phi $, similar to the calculation
of the charge and  spin currents. Accordingly, we find
\begin{equation}
\label{41}
{\bf S}(0)=\frac{e\, \Delta \phi }{2\pi \hbar }\left(
\frac{\tilde{\bf S}_\uparrow (0)}{v_\uparrow }
+\frac{\tilde{\bf S}_\downarrow (0)}{v_\downarrow }\right) ,
\end{equation}
where $\tilde{\bf S}_{\uparrow ,\downarrow }(0)$ can be found using
Eqs.~(36)-(38) with $n_x=0$ and the substitution $g_0\to \tilde{g}_0$,
which corresponds
to the scattering from the effective magnetic moment ${\bf M}_{eff}$.

Finally, we find the torque acting on a single localized moment in
the domain wall. For this purpose we use Eq.~(16) with ${\bf
S}(0)$ from (41), describing the spin accumulation created by
scattering from the domain wall as a whole. In its turn,
$\tilde{\bf S}_{\uparrow , \downarrow }(0)$ is calculated as
explained after Eq.~(41) using (36)-(38). The result can be
presented in the general  form
\begin{equation}
\label{42}
{\bf T}=\frac{j_0}{e}\left[
\eta \; {\bf n}\times \left( {\bf n}\times {\bf s}\right)
+\zeta \; {\bf n}\times {\bf s}\, \right] .
\end{equation}
where
\begin{equation}
\label{43}
\eta =\frac{g_0\, \tilde{g}_0\, (k_\downarrow ^2-k_\uparrow ^2)}
{2k_\uparrow k_\downarrow (k_\uparrow +k_\downarrow )^2
+\tilde{g}_0^2(k_\uparrow ^2+k_\downarrow ^2)}\; ,
\end{equation}
\begin{equation}
\label{44}
\zeta =-\frac{g_0\, (k_\uparrow +k_\downarrow )^2\,
[(k_\uparrow +k_\downarrow )^2-\tilde{g}_0^2]}
{2\left[ 2k_\uparrow k_\downarrow (k_\uparrow +k_\downarrow )^2
+\tilde{g}_0^2(k_\uparrow ^2+k_\downarrow ^2)\right] }\; ,
\end{equation}
and ${\bf s}$ is the unit vector along the spin polarization
corresponding to magnetization ${\bf M}$
at $x<-w$. The dependence of the coefficients $\eta $ and $\zeta $
on the parameters $\tilde{g}_0$ and on the electron gas
polarization $P=(k_\downarrow -k_\uparrow )/(k_\uparrow
+k_\downarrow )$ is presented in Figs.~2 and 3. As we see, both
coefficients strongly depend on the parameters of the ferromagnet
and on the parameters of the wall. In the case of small coupling
$\tilde{g}_0$, we obtain $\zeta \gg \eta $, i.e., the torque is
mostly related to the second component in Eq.~(42). In contrast,
if  $\tilde{g}_0$ is larger, the first term in (42) dominates.

\begin{figure}
\vspace*{-1cm}
\hspace*{-0.5cm}
\includegraphics[scale=0.45]{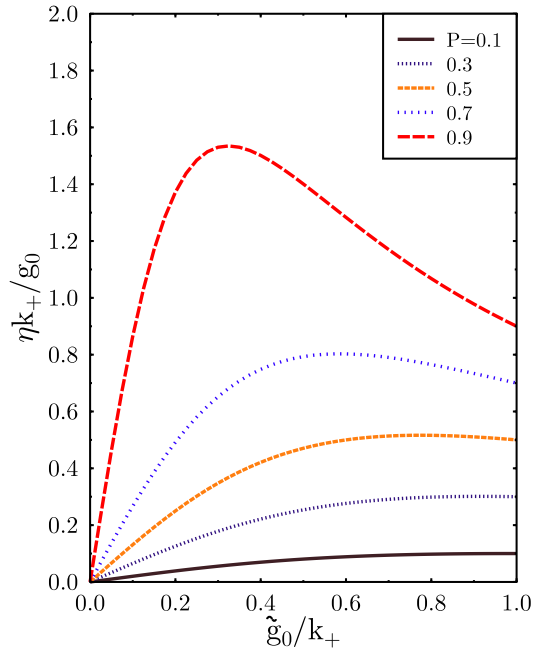}
\caption{Dependence of the factor $\eta $ on the effective
coupling $\tilde{g}_0$ for different values of the electron
polarization $P$.}
\end{figure}

\begin{figure}
\vspace*{-1cm}
\hspace*{-0.5cm}
\includegraphics[scale=0.45]{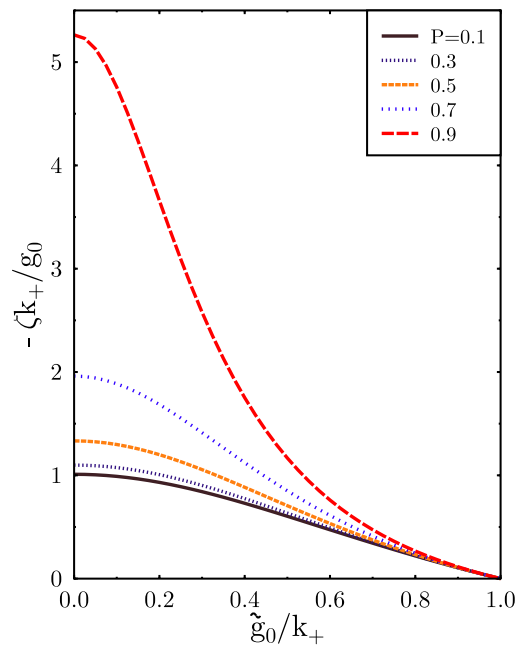}
\caption{Coefficient $\zeta $ vs. coupling constant $\tilde{g}_0$
for different values of $P$.}
\end{figure}

\subsection{Spin torque in $p$-type magnetic semiconductors}

The conductivity of magnetic semiconductors like GaMnAs is usually of
the $p$-type. The valence band of these semiconductors can be
described by a matrix Hamiltonian, which includes the spin-orbit
interaction.\cite{bir} Thus, the calculation of the transmission
of holes through the domain wall in GaMnAs semiconductors needs a
different model, which takes into account the complex band
structure.

Here we use the Luttinger model for the energy spectrum of holes
with the momentum $J=3/2$.\cite{luttinger56} We also neglect the
anisotropy of the energy spectrum. To simplify the calculations,
in this section we take the quantization axis $z$ along the wire.
Then in the quasi-one-dimensional case with the domain wall in the
$y-z$ plane, the Hamiltonian of holes is
\begin{eqnarray}
\label{45}
H=\frac{\hbar ^2}{2m_0}
\left( \gamma _1 +\frac{5\gamma _2}2\right)\frac{d^2}{dz^2}
-\frac{\hbar ^2\gamma _2}{m_0}\; J_z^2\; \frac{d^2}{dz^2}
\nonumber \\
-g\left[ J_yM_y(z)+J_zM_z(z)\right] ,
\end{eqnarray}
where $m_0$ is the mass of free electron, $\gamma _1$ and $\gamma
_2$ are the Luttinger parameters, and $J_\mu $ are the matrices of
the total momentum 3/2. Note that we are using Hamiltonian (45) to
describe the holes like unfilled electron states in the valence
band. Then the correct statistics of holes corresponds to the
reversed energy axis of holes as compared to that of electrons.

Similar to the previous consideration, we take the magnetization
$\bf M$ along the axis $z$ for $z<-w$ and in the opposite
direction for $z>w$, whereas in the region of $-w<z<w$ the moment
changes its orientation rotating in $y-z$ plane. For $z<-w$ the
holes can be described by the energy spectrum consisting of four
parabolic bands of the particles labelled by the momentum
projection $J_z$
\begin{equation}
\label{46}
E_{\pm 3/2}(k)=-\frac{\hbar ^2k^2}{2m_t}\mp \frac{3gM}2\; ,
\end{equation}
\begin{equation}
\label{47}
E_{\pm 1/2}(k)=-\frac{\hbar ^2k^2}{2m_l}\mp \frac{gM}2\; ,
\end{equation}
where $m_t=m_0/(\gamma _1-2\gamma _2)$ and $m_l=m_0/(\gamma
_1+2\gamma _2)$ are the masses of heavy and light holes,
respectively. In accordance with Eqs.~(46) and (47), the energy
band of the heavy holes with the moment projection $J_z=-3/2$ is
above all other bands. In the region of $z>0$ the spectrum is the
same but with the opposite signs  of $J_z$.

We assume that the holes are fully polarized so that the hole
density is rather small. Correspondingly, we assume that the
chemical potential $\mu $ is located in the interval of energies
$gM/2<\mu <3gM/2$.

The scattering state of holes corresponding to the wave $J_z=-3/2$
incoming from $z=-\infty $ is
\begin{eqnarray}
\label{48}
\psi ^\dag (z) =
\left( r_3^*e^{\kappa _3z},\;
r_2^*e^{\kappa _2z},\;
r_1^*e^{\kappa _1z},\;
e^{-ikz}+r^*e^{ikz}\right) ,
\nonumber \\
z<-w,
\end{eqnarray}
\begin{eqnarray}
\label{49}
\psi ^\dag (z) =
\left( t^*\, e^{-ikz},\;
t_1^*e^{-\kappa _1z},\;
t_2^*e^{-\kappa _2z},\;
t_3^*e^{-\kappa _3z}\right) ,
\nonumber \\
z>+w,
\end{eqnarray}
where $r,\ r_1,\, ... r_3$ and $t,\, t_1,\, ... t_3$ are the
reflection and the transmission coefficients, respectively. The
momentum of heavy hole $k$ is taken at the Fermi surface, $-\hbar
^2k^2/2m_t+3gM/2=\varepsilon$. The other momenta $\kappa _i$
correspond to the decaying components of the wave function,
$\kappa _1=\left[ 2m_l\left( \varepsilon -gM/2\right) /\hbar
^2\right] ^{1/2}$, $\kappa _2=\left[ 2m_l\left( \varepsilon
+gM/2\right) /\hbar ^2\right] ^{1/2}$, and $\kappa _3=\left[
2m_t\left( \varepsilon +3gM/2\right) /\hbar ^2\right] ^{1/2}$.
Note that the transmission coefficient $t$ in this notation
corresponds to the transmission from the state with moment
$J_z=-3/2$ to the state $J_z=3/2$.

In the limit of $w\to 0$, the matching condition can be presented
in the matrix form
\begin{eqnarray}
\label{50}
{\rm diag}\left(
m_t^{-1},\, m_l^{-1},\, m_l^{-1},\, m_t^{-1}\right)
\left( \left. \frac{d\psi }{dz}\right| _\delta
-\left. \frac{d\psi }{dz}\right| _{-\delta }\right)
\nonumber \\
-\lambda _0\, J_y\, \psi (0)=0,\hskip0.5cm
\end{eqnarray}
where $\lambda _0 =2gM_{eff}/\hbar ^2$.

Using Eq.~(50) and the continuity of the wave function at $z=0$ we
can calculate eight reflection and transmission coefficients. The
accumulated spin density ${\bf S}(0)$ induced by the current
flowing along $z$ can be calculated like in the previous section,
but with the opposite sign because the accumulation of polarized
holes means a loss of real particles (electrons). Then we find
\begin{equation}
\label{51}
S_x(0)=-\frac{e\, \Delta \phi }{2\pi \hbar v_t}\, {\rm Im}
\left( \sqrt{3}\, t_1^*t+2t_2^*t_1+\sqrt{3}\, t_3^*t_2\right) ,
\end{equation}
\begin{equation}
\label{52}
S_y(0)=-\frac{e\, \Delta \phi }{2\pi \hbar v_t}\, {\rm Re}
\left( \sqrt{3}\, t_1^*t+2t_2^*t_1+\sqrt{3}\, t_3^*t_2\right) ,
\end{equation}
\begin{equation}
\label{53}
S_z(0)=-\frac{e\, \Delta \phi }{4\pi \hbar v_t}
\left( 3\left| t\right| ^2+\left| t_1\right| ^2
-\left| t_2\right| ^2-3\left| t_3\right| ^2\right) ,
\end{equation}
where $v_t=\hbar k/m_t$ is the velocity of heavy holes at the Fermi level,
$e\, \Delta \phi =\varepsilon _{FR}-\varepsilon _{FL}>0$, and $\varepsilon _{FL}$
and $\varepsilon _{FR}$ are the Fermi levels at $z<-w$ and $z>w$, respectively.

\begin{figure}
\vspace*{-1cm} \hspace*{-0.5cm}
\includegraphics[scale=0.45]{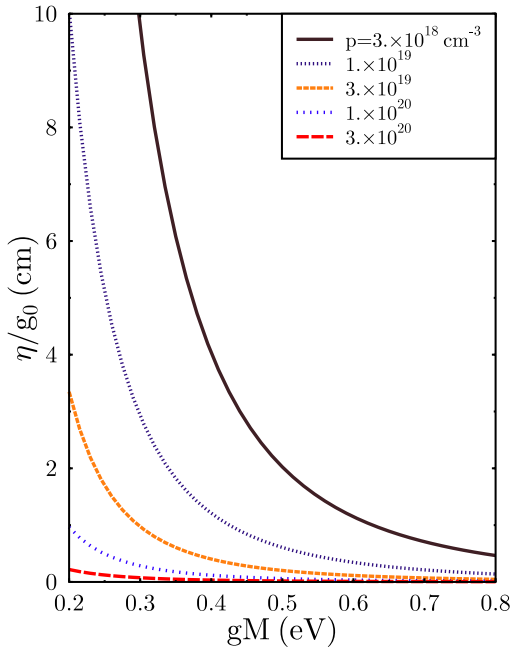}
\caption{Dependence of the factor $\eta $ on the magnetic
splitting $gM$ in the valence band of magnetic semiconductor for
different values of the bulk hole density $p$.}
\end{figure}

\begin{figure}
\vspace*{-1cm} \hspace*{-0.5cm}
\includegraphics[scale=0.45]{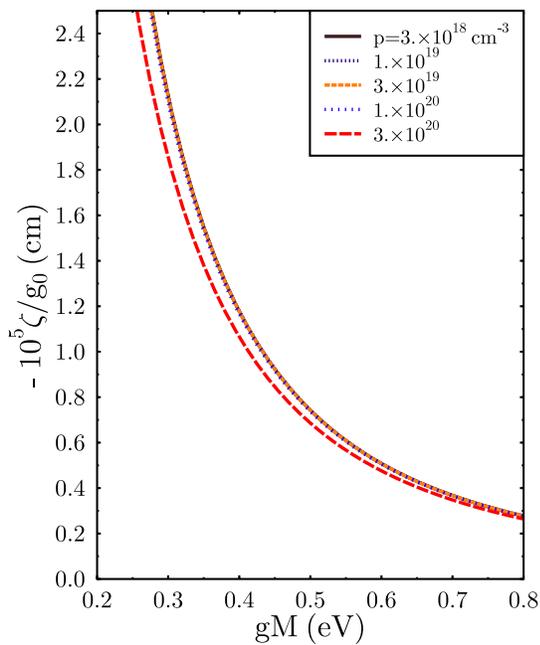}
\caption{Coefficient $\zeta $ vs magnetic splitting $gM$ for
different values of $p$.}
\end{figure}

Using Eqs.~(16) and (42), we find the parameters $\eta $ and $\zeta $
determining the torque acting on a single magnetic moment $M_0$
\begin{equation}
\label{54}
\eta =\frac{egM_0}{j_0\hbar }\; S_z(0),
\end{equation}
\begin{equation}
\label{55}
\zeta =-\frac{egM_0}{j_0\hbar }\; S_x(0),
\end{equation}
where $j_0=-e^2\, \Delta \phi \left| t\right| ^2/2\pi \hbar $, and
the "-" sign in current is due to the opposite charge of holes.

The dependence of $\eta $ and $\zeta $ on the magnitude of magnetic splitting
$gM$ for different bulk hole densities $p$ is presented in Figs.~4 and 5.
We take the cross section $A=1$~nm$^2$, and the momentum of heavy holes
$k=\pi p_{1D}$, where $p_{1D}$ is the linear density of holes.

As we can see from Figs.~4 and 5, the factor $\zeta $ is
negligibly small as compared to $\eta $. In our model, the density
of holes and the spin splitting are independent parameters. Thus,
the magnitude of torque $\eta $ increases with the decreasing hole
density $p$ at a fixed value of $gM$. However, in real magnetic
semiconductors these values are not independent, and the magnetic
splitting increases with the increasing hole
density.\cite{dietl01}

It should be noted that the used condition of $w\to 0$ implies
that not only the wavelength of holes with $J_z=3/2$ is large as
compared to the domain wall width, $kw\ll 1$, but also the
conditions $\kappa _iw\ll 1$ for all $\kappa _i$ should be
fulfilled. This condition is restrictive for the magnitude of the
magnetic splitting, $(gMm_t)^{1/2}w/\hbar \ll 1$.

\section{Motion of the domain wall}

\subsection{Hamiltonian and equations of motion}

Now we consider the Hamiltonian $\mathcal{H}_0$ describing a
quasi-one-dimensional magnetic system with a domain wall. We adopt
a model including the magnetic exchange interaction and two
different anisotropy constants $\lambda _1$ and $\lambda _2$ in
the $z$ and $y$ directions, respectively (see Fig.~1)
\begin{equation}
\label{56}
\mathcal{H}_0
=\frac{a}2 \left( \frac{\partial {\bf n}}{\partial x}\right) ^2
+\frac{\lambda _1}2\, n_z^2+\frac{\lambda _2}2\, n_y^2\, ,
\end{equation}
where $a$ is the constant of exchange interaction, and ${\bf
n}(x)$ is the unit vector of magnetization. Using this Hamiltonian
we are going to describe the magnetic nanowire like presented in
Fig.~1. Correspondingly, we assume that the magnetization vector
field ${\bf n}$ depends only on the coordinate $x$ and time $t$.
The Hamiltonian $\mathcal{H}_0$ is the energy density of the
magnetic system in the absence of the spin torque.

In the following we restrict ourselves by considering the
above-calculated spin torque as a driving force acting on the
domain wall. Hence, we neglect the direct transfer of momentum
from electrons reflected from the domain wall. As we show in
Appendix, this effect is much smaller than the above-calculated
spin torque. On the other hand, our consideration has a general
character without specifying the mechanism determining the values
of the factors $\eta $ and $\zeta $.

Using the spherical angles $\theta (x,t)$ and $\varphi (x,t)$, we
can rewrite the Hamiltonian $\mathcal{H}_0$ as
\begin{eqnarray}
\label{57}
\mathcal{H}_0=\frac{a}2\left( \frac{\partial \theta }{\partial x}\right) ^2
+\frac{a}2\left( \frac{\partial \varphi }{\partial x}\right) ^2\sin ^2 \theta
+\frac{\lambda _1}2\cos ^2\theta
\nonumber \\
+\, \frac{\lambda _2}2\sin ^2\theta \, \sin^2\varphi .
\end{eqnarray}

The Landau-Lifshitz-Gilbert equation of motion includes a damping
term and two possible components of the current-induced torque, as
discussed in the previous section,
\begin{eqnarray}
\label{58}
\frac1{\gamma }\; \frac{\partial {\bf n}}{\partial t}
=-{\bf n}\times \left( \frac{\delta \mathcal{H}_0}{\delta {\bf n}}
-\frac{\partial }{\partial x}
\frac{\delta \mathcal{H}_0}{\delta (\partial {\bf n}/\partial x)}\right)
%\nonumber \\
-\alpha \, {\bf n}\times \frac{\partial {\bf n}}{\partial t}
\nonumber \\
+J_0\zeta \, {\bf n}\times {\bf s} +J_0\eta \, {\bf n}\times
\left( {\bf n}\times {\bf s}\right) ,\hskip0.5cm
\end{eqnarray}
where $\alpha $ is the damping constant, $\gamma =g\mu _B/\hbar M$
is the gyromagnetic ratio divided by $M$, $J_0=j_0\hbar /eg\Omega _0$, and
$\Omega _0$ is a volume per magnetic moment. In Eq.~(58) the
spin torque is expressed in terms of transferred moment per
unit volume, and it enters directly the equation of motion. The
corresponding spin-torque terms in the magnetic Hamiltonian can be
presented as
\begin{equation}
\label{59}
\mathcal{H}_{int}=J_0\zeta \; {\bf n}\cdot {\bf s}
+J_0\eta \int _0^1d\tau \;
{\bf n}\cdot \left( \frac{\partial {\bf n}}{\partial \tau }
\times {\bf s}\right) ,
\end{equation}
where ${\bf n}(\tau =0)=0$ and ${\bf n}(\tau =1)={\bf n}$.

The Lagrangian of the magnetic system contains a term with a time
derivative as follows\cite{braun96,takagi96}
\begin{equation}
\label{60}
\mathcal{L}=A\int dx\left[
\frac1{\gamma }\, \frac{\partial \varphi }{\partial t}\;
\left( \cos \theta -1\right) -\mathcal{H}\right] .
\end{equation}
Neglecting the damping term, the Landau-Lifshitz equations of the
magnetic dynamics take the form
\begin{eqnarray}
\label{61}
\frac1{\gamma }\; \frac{\partial \theta }{\partial t}
=-a\; \frac{\partial ^2\varphi }{\partial x^2}\; \sin \theta
+\lambda _2\, \sin \theta \, \sin \varphi \, \cos \varphi
\nonumber \\
-J_0\eta \, \cos \theta \, \cos \varphi ,
-J_0\zeta \, \sin \varphi ,
\end{eqnarray}
\begin{eqnarray}
\label{62}
\frac{\sin \theta }{\gamma }\; \frac{\partial \varphi }{\partial t}
=a\; \frac{\partial ^2\theta }{\partial x^2}
-a \left( \frac{\partial \varphi }{\partial x}\right) ^2
\sin \theta\, \cos \theta
\nonumber \\
+\lambda _1\, \cos \theta \, \sin \theta
-\lambda _2\, \sin \theta \, \cos \theta \, \sin ^2\varphi
\nonumber \\
+J_0\eta \, \sin \varphi
+J_0\zeta \, \cos \theta \, \cos \varphi .
\end{eqnarray}
In the absence of current, $j_0=0$, they have the
well-known\cite{landau8} kink-like static solution $\varphi
_0(x)=\arccos \left[ \tanh ( \beta _0x)\right] $ and $\theta
_0=\pi /2$, where $\beta _0=\left( \lambda _2/a\right) ^{1/2}$ is
the inverse width of the static domain wall. From now on we assume
for definiteness that $\lambda _1>\lambda _2$, so that the static
domain wall with the magnetization in the $x-y$ plane is
energetically more favorable.

In a general case, the solution of the nonlinear dynamical
Eqs.~(61)-(62), describing the moving domain wall, is a difficult
problem. Therefore, we assume in the following that one of the
anisotropy constants is large, $\lambda _1\gg \lambda _2$.

\subsection{Strong easy-plane anisotropy}

We consider the case of a large easy-plane anisotropy, and,
accordingly, assume that for the moving domain wall (subjected to
the torque) the deviation of magnetization vector ${\bf M}$ from
the $x-y$ plane is small. Then we can write $\theta (x,t)=\pi
/2+\chi (x,t)$ and take $|\chi (x,t)|\ll 1$. Up to the second
order in $\chi (x,t)$ field, the Lagrangian $\mathcal{L}$ is
\begin{eqnarray}
\label{63}
\mathcal{L}=A\int dx\left[
-\frac1{\gamma }\, \frac{\partial \varphi }{\partial t}
\left( \chi +1\right)
-\frac{a}2 \left( \frac{\partial \chi }{\partial x}\right) ^2
-\frac{a}2\left( \frac{\partial \varphi }{\partial x}\right) ^2
\right. \nonumber \\ \left.
\times \left( 1-\chi ^2\right)
-\frac{\lambda _1}2\, \chi ^2
-\frac{\lambda _2}2\, \sin ^2\varphi \left( 1-\chi ^2\right)
%\right. \nonumber \\ \left.
+J_0\eta \; \chi \sin \varphi
\right. \nonumber \\ \left.
+J_0\zeta \; \cos \varphi
\right] ,\hskip0.3cm
\end{eqnarray}

Since we restricted the treatment to quadratic terms in $\chi $ in
the Lagrangian, the integral over $\chi $ is Gaussian, and we can
integrate out\cite{feynman} the $\chi $ fields to obtain
\begin{eqnarray}
\label{64}
\mathcal{L}=A\int dx\left[
\frac12\int dx^\prime \;
G(x,x^\prime )
\left(
\frac1{\gamma } \frac{\partial \varphi (x)}{\partial t}
-J_0\eta \, \sin \varphi (x)\right)
\right. \nonumber \\ \left.
\times \left(
\frac1{\gamma }\frac{\partial \varphi (x)}{\partial t}
-J_0\eta \, \sin \varphi (x)\right)
-\frac{a}2\left( \frac{\partial \varphi }{\partial x}\right) ^2
%\right. \nonumber \\ \left.
-\frac{\lambda _2}2\, \sin ^2\varphi
\right. \nonumber \\ \left.
+J_0\zeta \, \cos \varphi
\right] ,\hskip0.5cm
\end{eqnarray}
where the Green function $G(x,x)$ obeys the following
equation
\begin{eqnarray}
\label{65}
\left[ -a\; \frac{\partial ^2}{\partial x^2}
-a\left( \frac{\partial \varphi }{\partial x}\right) ^2
+\lambda _1 -\lambda _2\sin ^2\varphi \right]
G(x,x^\prime )
\nonumber \\
=\delta (x-x^\prime ) .
\end{eqnarray}
Note that $\varphi $-fields are taken at the same time $t$ in
Eq.~(64). It follows from the equation for Green function
$G(x,t;\, x^\prime ,t^\prime )\sim \delta (t-t^\prime )$
describing the propagation in time.

Equation (64) contains
$\varphi (x,t)$, which should be the saddle point solution of the
Lagrangian, i.e., the self-consistency should be preserved.

Neglecting the first term in Eq.~(65) we can find an approximate
formula for the Green function proportional to $\delta (x-x^\prime )$
\begin{equation}
\label{65}
G(x,x^\prime )=\delta (x-x^\prime )
\left[ -a\left( \frac{\partial \varphi }{\partial x}\right) ^2
+\lambda _1 -\lambda _2\sin ^2\varphi \right] ^{-1}.
\end{equation}
This form of $G(x,x^\prime )$ leads to the point interaction
of the $\varphi $-fields in
the first term of Eq.~(64). Physically, neglecting the first
term with derivatives in Eq.~(65), we substitute the finite-range
interaction by the $\delta $-like one.

One can estimate at which conditions the use the
Green function in the form of Eq.~(66) is justified.
The exact solution of
Eq.~(65) can be presented as
\begin{equation}
\label{67}
G(x,x^\prime )
=\sum _n\frac{\phi _n(x)\, \phi _n^*(x^\prime )}
{\varepsilon _n+\lambda _1}\; ,
\end{equation}
where $\phi _n(x)$ and $\varepsilon _n$ are the eigenfunctions
and corresponding eigenvalues of the equation
\begin{equation}
\label{68}
\left[ -a\; \frac{\partial ^2}{\partial x^2}
-a\left( \frac{\partial \varphi }{\partial x}\right) ^2
-\lambda _2\sin ^2\varphi -\varepsilon _n\right] \phi _n(x)=0.
\end{equation}
We expect that the function $\varphi (x)$ in Eqs.~(65) and (68) is
similar to the form of the static solution $\varphi _0(x)$. Thus,
Eq.~(68) corresponds to the Schr\"odinger equation for a particle
of mass $m=\hbar ^2/2a$ in the potential well $V(x)$ of width
$L_0\sim (a/\lambda _2)^{1/2}$. The energy spectrum of this
problem consists of a level in the well $\varepsilon _0\simeq
-\lambda _2$ and a continuous spectrum for all positive energies.

For $\varphi (x)=\varphi _0(x)$, the potential has the form of
$V(x)=-2\lambda _2/\cosh ^2(\beta _0x)$, and the discrete energy
spectrum\cite{landau3} has one level $\varepsilon _0=-4\lambda
_2$. The eigenfunctions $\phi _n(x)$ corresponding to the
continuous spectrum, are oscillatory functions, so that their
contribution to Eq.~(67) can be estimated as
$G^{(cont)}(x,x^\prime )\simeq (a\lambda _1)^{-1/2}e^{-\kappa
_1|x-x^\prime |}$, where $\kappa _1=(\lambda _1/a)^{1/2}$. Thanks
to $\lambda _1\gg \lambda _2$, it is a strongly localized function
on the scale of the distance (static domain wall width) $L_0$. On
the other hand, the contribution of the localized state gives us
$G^{(0)}(x,x^\prime )\simeq (1/L_0\lambda _1)\, e^{-\kappa
_0|x-x^\prime |}$, where $\kappa _0=1/L_0$. Thus, in the case of
$\lambda _1\gg \lambda _2$ (i.e., strong in-plane anisotropy), the
contribution of $G^{(0)}$ can be neglected as compared to the
short-range interaction. Using the condition of strong easy-plane
anisotropy, $\lambda _1\gg \lambda _2$, we can simplify Eq.~(66)
essentially and obtain the expression
\begin{equation}
\label{69}
G(x,x^\prime )\simeq \frac{\delta (x-x^\prime )}{\lambda _1}\; .
\end{equation}
In this approximation, the Lagrangian (64) acquires the following form
\begin{eqnarray}
\label{70}
\mathcal{L}=A\int dx\left[
\frac1{2\lambda _1}
\left(
\frac1{\gamma }\frac{\partial \varphi }{\partial t}
-J_0\eta \, \sin \varphi \right) ^2
%\right. \nonumber \\ \left.
%\times \left(
%\frac{\partial \varphi }{\partial t}
%-\frac{j_0\eta }{e}\sin \varphi \right)
-\frac{a}2\left( \frac{\partial \varphi }{\partial x}\right) ^2
\right. \nonumber \\ \left.
-\frac{\lambda _2}2\, \sin ^2\varphi
+J_0\zeta \, \cos \varphi
\right] .\hskip0.5cm
\end{eqnarray}
The corresponding saddle-point equation is
\begin{eqnarray}
\label{71}
-\frac1{\gamma ^2\lambda _1}\;
\frac{\partial ^2\varphi }{\partial t^2}
%-\frac{j_0\eta }{\gamma ^2\lambda _1e}\;
%\frac{\partial \varphi }{\partial t}\; \cos \varphi
+\frac{J_0^2\eta ^2}{\lambda _1}\;
\sin \varphi \, \cos \varphi
+a\; \frac{\partial ^2\varphi }{\partial x^2}
\nonumber \\
-\lambda _2\sin \varphi \, \cos \varphi
-J_0\zeta \; \sin \varphi =0.\hskip0.5cm
\end{eqnarray}
Now we can study the problem of the domain wall dynamics in terms
of a single $\varphi (x,t)$ field.

\subsection{Solution for $\zeta =0$}

Let us consider the possibility of the kink-like solution moving
with an arbitrary constant velocity $v$, $\varphi (x,t)\equiv
\varphi (x-vt)$. We can find such solutions in the case of $\zeta =0$,
trying a function which obeys the equality $\partial \varphi
(x)/\partial x=\beta \sin \varphi (x)$. This function differs from
the static solution only by a different choice of $\beta $ instead
of $\beta _0=(\lambda _2/a)^{1/2}$. Substituting it into (71), we
obtain the equation that relates the values of $\beta $ and $v$ as
\begin{equation}
\label{72}
\beta ^2\left( a-\frac{v^2}{\gamma ^2\lambda _1}\right)
%+\beta \; \frac{j_0\eta v }{\gamma ^2\lambda _1e}
-\lambda _2+\frac{J_0^2\eta ^2}{\lambda _1}=0.
\end{equation}
The dependence $\beta (\tilde{v})$ is presented in Fig.~6 where we denoted
$\tilde{v}=v/\gamma \sqrt{\lambda _1a}$ and
$\tilde{j}_0=J_0\eta /\sqrt{\lambda _1\lambda _2}$.

When $j_0=0$, we find from (72) that $\beta ^2=\beta
_0^2/(1-v^2/\gamma ^2\lambda _1a)$. It means, that in the absence
of the current, the solution for a moving domain wall is more
sharp as compared to the static wall.

\begin{figure}
\vspace*{-1cm}
\hspace*{-0.5cm}
\includegraphics[scale=0.45]{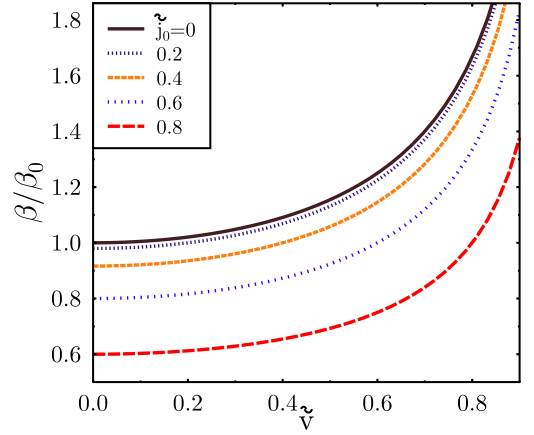}
\caption{Dependence of the parameter $\beta $ on the domain wall
velocity $\tilde{v}$ for different values of the current.}
\end{figure}

For $j_0\ne 0$ and $v\to 0$, the value of $\beta $ depends on the
current as
$\beta ^2=\beta _0^2\, (1-J_0^2\eta ^2a/\lambda _1\lambda _2)$,
i.e., the current makes the thickness of the static domain wall larger.

Now we can use the velocity $v$ as a variational parameter to minimize
the Lagrangian
\begin{equation}
\label{73}
\mathcal{L}=\frac{A\, F(v)}{2\beta (v)}
\int \sin ^2\varphi (x)\; d(\beta x),
\end{equation}
where
\begin{equation}
\label{74}
F(v)=\frac1{\lambda _1}
\left[ \frac{v\, \beta (v)}{\gamma }+J_0\eta \right] ^2
-a\beta ^2-\lambda _2,
\end{equation}
the function $\beta (v)$ is defined by Eq.~(72), and the integral in (73) does
not depend on $v$.

Using Eqs.~(72) to (74) we find that for $j_0=0$,
the quantity $F(v)=-\lambda _2$ for any $v$.
Thus, the minimum of $\mathcal{L}$ corresponds
to $\beta =\beta _0$, which is the minimum value of the dependence $\beta (v)$
for $j_0=0$. In the limit of a small velocity, $v^2\ll \gamma ^2\lambda _1a$,
and using the relation $\int \sin ^2\varphi (\alpha )\, d\alpha =2$,
we find the kinetic energy of the moving domain wall in the form of
$E_{kin}=m^*v^2/2$, where
$m^*=A\sqrt{\lambda _2}/\gamma ^2\lambda _1\sqrt{a}$ is the effective mass
of the domain wall. This is in agreement with the definition from
Ref.~[\onlinecite{malozemoff79}] for $\lambda _1=2\pi M^2$.

For $j_0\ne 0$, we can present the dependence of the factor $F$ on both parameters
$v$ and $j_0$ as
\begin{equation}
\label{75}
F(v)=-\lambda _2\left[ 1-2\tilde{j}_0^2-2\tilde{j}_0\, \tilde{v}
\left( \frac{1-\tilde{j}_0^2}{1-\tilde{v}^2}\right) ^{1/2}\right] .
\end{equation}
In the limit of $v\to 0$,
the factor $F$ changes its sign for $j_0>j_{0\, cr}$, where
\begin{equation}
\label{76}
j_{0\, cr}=\frac{e\Omega _0\sqrt{\lambda _1\lambda _2}}
{\sqrt{2}\, \hbar \eta }
\end{equation}
is the critical current.
Thus, if $j_0>j_{0\, cr}$, the solution with moving domain wall is
energetically favorable.

We can interpret the effect of the current as
leading to an effective reduction of the effective mass of the domain wall.
For $j_0>j_{0\, cr}$ the current induces an  instability towards
a spontaneous motion of the wall.

\subsection{Case of $\zeta \ne 0$}

In the case of $\zeta \ne 0$, there are no solutions of Eq.~(71)
corresponding to the motion of the domain wall with a constant velocity
because the last term in this equation acts as a force accelerating
the domain wall. Indeed, if we assume a probe solution in the form of
$\varphi (x,t)\equiv \varphi [x-x_0(t)]$
we find
\begin{equation}
\label{77}
m^*\, \ddot{x}_0(t)+J_0A\zeta =0,
\end{equation}
where $m$ is a constant in the limit of a  small velocity.
In other words, Eq.~(77) describes the acceleration of the
domain wall just after we apply some voltage.
Hence, our model can describe the steady
state if we include a viscosity (friction) into the equation of motion.
We can use the damping term from Eq.~(58).
Writing the
corresponding additional term in Eq.~(71) as
$F_d=-\alpha \, \partial \varphi /\partial t$, we find
the following equation that determines the velocity of the moving
domain wall as follows
\begin{equation}
\label{78} v\, \beta (v)\simeq \frac{J_0\zeta }{\alpha }.
\end{equation}
This equation indicates  a linear dependence of the velocity on the current
in the limit of a small velocity, when $\beta $ is constant.
As we see from Eq.~(78), it corresponds to a large damping.

Effective friction may as well stem from the
pinning by impurities.
This case  can be described phenomenologically leading to
another mechanism for the critical current.\cite{takagi96}

\section{Conclusions}

We calculated the components of the spin torque acting on a thin
domain wall in a magnetic nanowire subject to an electric current.
These components can induce a rotation of magnetic moments in
different directions.

We also considered the dynamics of a domain wall in the presence
of the current. We demonstrated that a moving magnetic kink,
similar to the static domain wall, can be a solution of the
equations for the  magnetic dynamics only at some special
conditions. We found these conditions in the case of a large ratio
of the magnetic anisotropy constants. In the limit of small
velocities, the solution does look like a kink but its width
decreases with increasing velocity. In the limit of a small
velocity, the domain wall moves like a particle of a mass
determined by the exchange interaction and anisotropies. One of
the spin torque components $\zeta $, dominating at the small
coupling, acts as a moving force on the domain wall, accelerating
the wall, provided that there is no pinning to impurities.

Recent direct observations of the domain-wall configurations show
that the spin structure of the wall changes with the current, and
the structure depends on the velocity of the domain wall
motion.\cite{klaui05}

We performed the calculation of torque in the limit of thin domain
wall, $w\ll \lambda _F$. It allows to simplify essentially the
problem and to do all the calculations analytically. In reality,
this inequality may be not well satisfied even in the magnetic
semiconductors. Here we present the estimations for the case when
the above-mentioned inequality is obeyed.

Let us take the cross section of the wire $A=1$~nm$^2$ and the
bulk carrier density $n_{3D}=10^{19}$~cm$^{-3}$, corresponding to
the linear density $n_{1D}=n_{3D}A=10^5$~cm$^{-1}$. It gives us
$k_F=\pi n_{1D}\simeq 3\times 10^5$~cm$^{-1}$, and the carrier
wavelength $\lambda _F=2\pi /k_F\simeq 100$~nm.

To estimate the domain wall width, we assume the magnitude of
magnetization $M=100$~Oe, the demagnetizing factor along the $y$
axis $n^{(y)}=0.3$, and calculate the anisotropy constant as
$\lambda _2\simeq 8\pi n^{(y)}\, M^2\simeq 10^{5}$~erg/cm$^3$. For
the energy of magnetic interaction $E_{int}\simeq 10$~meV at a
distance between magnetic ions of $c_0=1$~nm, the exchange
parameter $a=E_{int}\, c_0/A\simeq 10^{-8}$~erg/cm. Then the domain
wall width has a reasonable value of $w=(a/\lambda _2)^{1/2}\simeq
10$~nm. Comparing these estimations, we see that the main
inequality of $w\ll \lambda _F$ is satisfied. At a larger carrier
density, both $w$ and $\lambda $ can be of the same order of
magnitude or the inequality is reversed like in magnetic metals.
In this case, the constants $\zeta $ and $\eta $ should be
calculated numerically.

\begin{acknowledgments}
V.D. thanks P. M. Levy for valuable discussions concerning the accumulation of spin.
This work is supported by FCT Grant No.~POCI/FIS/58746/2004
(Portugal) and by Polish State Committee for Scientific
Research under Grants Nos.~PBZ/KBN/044/P03/2001 and 2~P03B~053~25.
One of the authors (V.D.) thanks the Calouste Gulbenkian Foundation
in Portugal for support.
\end{acknowledgments}

\appendix

\section{Spin torque due to the momentum transfer}

The reflection of electrons from the domain wall is accompanied by
the transfer of momentum from the electron system to the domain
wall. In the presence of electric current transmitted through the magnetic
wire, it creates an additional force acting on the
wall.\cite{berger92} Here we estimate the magnitude of this effect
in the case of a thin DW, $k_Fw\ll 1$.

The force $F$ is determined by the total transferred momentum in
unit time. Taking into account the contribution of spin up and
down scattering states corresponding to the waves incoming from
$-\infty $  in the energy range between $\varepsilon _F$ and
$\varepsilon +e\, \Delta \phi $, we find
\begin{eqnarray}
\label{A1}
\mathcal{F}=\frac{e\Delta \phi }{2\pi }\left[
k_\uparrow \left(
1+\left| r_\uparrow \right| ^2
-\left| t_{\uparrow f}\right| ^2
+\frac{v_\uparrow }{v_\downarrow }\left| r_{\downarrow f}\right| ^2
-\frac{v_\uparrow }{v_\downarrow }\left| t_\downarrow \right| ^2\right)
\right. \nonumber \\ \left.
+k_\downarrow \left(
1+\left| r_\downarrow \right| ^2
-\left| t_{\downarrow f}\right| ^2
+\frac{v_\downarrow }{v_\uparrow }\left| r_{\uparrow f}\right| ^2
-\frac{v_\downarrow }{v_\uparrow }\left| t_\uparrow \right| ^2\right)
\right] .\hskip0.3cm
\end{eqnarray}
This force tends to shift the domain wall along the $x$ direction.
For a local moment within the wall it is equivalent to the appearance
of a torque. To estimate the magnitude of this mechanical torque
acting on a single moment we use a simplified model.

We describe
the domain wall by the $\varphi (x)$ field, which is the angle in
$x-y$ plane determining the direction of moment ${\bf
M}(x)$ as shown in Fig.~1. We assume that the shift along $x$ is
related to the following interaction
\begin{equation}
\label{A2}
\mathcal{H}_{int}=\lambda \varphi (x) \, v(x),
\end{equation}
where $\lambda $ is a constant,
$v(x)=-d\varphi _0/dx$, and $\varphi _0(x)$ is the
static solution for the domain wall. The potential $v(x)$ has the
form of a potential well in the vicinity of the domain wall, and
it forces (makes energetically favorable) a correction to the
$\varphi (x)$ field of the same form, $\delta \varphi (x) \sim
d\varphi _0/dx$. On the other hand, the correction $\delta \varphi
(x)=\left( d\varphi /dx\right) \delta x_0$ is the shift along
the axis $x$ by $\delta x_0$. Thus, the interaction term in form
of (A2) in the equation of motion for the $\varphi (x)$ acts as a
shifting force.

The constant $\lambda $ should be determined by the condition that
the energy $\delta E$ associated with the shift, gives the force
$\mathcal{F}$:
\begin{equation}
\label{A3}
\mathcal{F}=-\frac{\delta E}{\delta x_0}=\lambda A\int
\left( \frac{d\varphi _0}{dx}\right) ^2dx.
\end{equation}
Using the known solution, $d\varphi _0/dx=\beta \sin \varphi _0(x)$,
we find $\lambda =\mathcal{F}/2\beta A$.

The equation of motion for $\varphi (x)$ (64) includes the
additional torque term as $\lambda v(x)=\mathcal{F}v(x)/2\beta A$.
Using (A1) we estimate the torque acting on the localized moment
$M_0=M\Omega $
\begin{equation}
\label{A4}
T_{mt}\simeq \frac{j_0}{e}\; \frac{k_F\Omega }{A}.
\end{equation}
where $\Omega $ is the volume of an elementary cell.
We find that the relative contribution of the
momentum-induced torque with respect to the spin transfer is
\begin{equation}
T_{mt}/T_{st}\simeq k_F\Omega /A\ll 1.
\end{equation}

\end{document}